\input{aipcheck}

\documentclass[
    ,final            
  ]
  {aipproc}

\layoutstyle{6x9}

\begin{document}

\title{Experimental status of deeply bound kaonic states in nuclei}


\author{V.K. Magas}{
  address={Departament d'Estructura i Constituents de la Materia, Universitat de Barcelona, Diagonal 647, E-08028 Barcelona, Spain}
}

\author{J. Yamagata-Sekihara}{
  address={Yukawa Institute for Theoretical Physics, Kyoto University, Kyoto 606-8502, Japan},
  ,altaddress={Departamento de F\'{\i}sica Te\'orica and IFIC, Centro Mixto Universidad de Valencia-CSIC,
   Institutos de Investigaci\'on de Paterna, Apartado 22085, 46071 Valencia, Spain} 
}

\author{S. Hirenzaki}{
  address={Department of Physics, Nara Women's University, Nara 630-8506, Japan}
}

\author{E. Oset}{
  address={Departamento de F\'{\i}sica Te\'orica and IFIC, Centro Mixto Universidad de Valencia-CSIC,
   Institutos de Investigaci\'on de Paterna, Apartado 22085, 46071 Valencia, Spain}
}

\author{A. Ramos}{
  address={Departament d'Estructura i Constituents de la Materia, Universitat de Barcelona, Diagonal 647, E-08028 Barcelona, Spain}
}

\begin{abstract}
 We review recent claims of the existence of deeply bound 
kaonic states in nuclei. Also we study in details the $(K^-,p)$ reaction on $ ^{12}C$ with $1$ GeV/c momentum kaon beam, based on which 
a deep kaon nucleus optical potential was claimed in \cite{Kishimoto:2007zz}. In our Monte Carlo simulation of this reaction we include not only the quasi-elastic $K^-p$ scattering, as in \cite{Kishimoto:2007zz}, but also 
$K^-$ absorption by one and two nucleons followed by the decay of the hyperon in $\pi N$, 
which can also produce strength in the region of interest. The final state interactions in terms of multiple scattering of the 
$K^-$, $p$ and all other primary particles on their way out of the nucleus is also considered.
We will show that all these additional mechanisms allow us to explain the observed spectrum with a "standard"
shallow kaon nucleus optical potential obtained in chiral models.   

\end{abstract}

\maketitle


    The issue of the kaon interaction in the nucleus has attracted much
attention in past years. Although from the study of kaon atoms one knows that
the $K^-$-nucleus potential is attractive \cite{friedman-gal}, the discussion
centers on how attractive the potential is and whether it can accommodate deeply
bound kaon atoms (kaonic nuclei), which could be observed in direct reactions.

All modern potentials
based on underlying chiral dynamics of the $KN$ interaction
\cite{lutz,angelsself,schaffner,galself,Tolos:2006ny} lead to
moderate potentials of the order of 60 MeV attraction at normal nuclear density.
They also have a large imaginary part making the width of the bound states
much larger than the energy separation between the levels, which would rule out
the  experimental observation of these states.

Deep  $K^-$-N optical potentials are preferred by the phenomenological fits to kaon atoms data. 
One of the most known extreme cases of this type is a highly
attractive phenomenological potential with about 600 MeV strength in the center of the nucleus, 
introduced in \cite{akaishi:2002bg,akainew}.  
In this picture  such an attractive $K^-$, inserted inside the nucleus,  would lead to a shrinkage of the nucleus, generating a new very compact object - kaonic nucleus - with a central density which can be 10 times larger than normal nuclear density. Such super-deep
potentials were criticized in \cite{toki,Hyodo:2007jq,Oset:2007vu,npangels}.

From the experimental side the search for deeply bound $K^-$ states with nucleons is a most direct and clear way to answer whether the $K^-$-nucleon potential is deep or shallow, because only a deep potential may generate states sufficiently narrow to be observed experimentally. Experimental attempts to resolve this situation have been made since 2004, but the situation is still very unclear. 

Several claims of observed deeply bound $K^-$ states have been made. However, the first one, $K^-pnn$ deeply bound state from the 
experiment at KEK \cite{Suzuki:2004ep}, is now withdrawn after a new more precise experiment \cite{Sato:2007sb}.

Two other claims  of the existence of deeply bound 
kaonic states in nuclei came from the observation by the FINUDA collaboration of some
peaks in the $(\Lambda p)$ \cite{Agnello:2005qj} and $(\Lambda d)$ \cite{:2007ph} 
invariant mass distributions, following the absorption of stopped K- in different nuclei.  
These were interpreted in terms of deeply bound $K^-pp$  and $K^-ppn$  clusters correspondingly. However, recently it has been shown that these peaks are naturally explained in terms of $K^-$ absorption by two  \cite{Magas:2006fn,Ramos:2007zz,Crimea} or 
three \cite{Magas:2008bp}  nucleons respectively, leaving the rest of the original nuclei as spectator. And for the reactions on heavy nuclei 
the subsequent interactions of the particles produced in the primary absorption process ($\Lambda$, $p$ etc.) 
with the residual nucleus have to be taken into account \cite{Magas:2006fn,Ramos:2007zz,Crimea}. 

There are also claims of $K^-pp$ and $K^-ppn$ bound states from $\bar{p}$ annihilation in $ ^4He$ at rest measured by OBELIX@CERN \cite{Obelix}, however their statistical significance is very low. The most recent is the claim of $K^-pp$ bound state, seen in $pp\rightarrow K^+ X$  reaction, from DISTO experiment \cite{DISTO}. These experimental claims are under investigation now.  Before calling in new physics one has to make sure that these data cannot be explained with conventional mechanisms. Also, it is worth mentioning that all above mentioned "experimental claims" are in disagreement with each other.  
 
There is, however, one more experiment where the authors claim
the evidence for a strong kaon-nucleons potential, with a depth
of the order of 200 MeV \cite{Kishimoto:2007zz}.  
The experiment
looks for fast protons
emitted from the absorption of in flight kaons by $^{12}C$ in coincidence with at least one charged particle in the decay counters sandwiching the target.
The data analysis in \cite{Kishimoto:2007zz} is based on the assumption that the coincidence requirement does not change the shape of the final spectra.
We shall see that this assumption doesn't hold and the
interpretation of the data requires a more thorough approach than the one
used in  that work. 

One of the shortcomings of Ref.~\cite{Kishimoto:2007zz} stems from employing the
Green's function method \cite{Morimatsu:1994sx} to analyze the data.  
The only mechanism considered in
Ref.~\cite{Kishimoto:2007zz} for the emission of fast protons is the $\bar{K} p \to
\bar{K} p$ process, taking into account the optical potential for the slow
kaon in the final state. 
We shall show that there are other mechanisms that
contribute to generate fast protons, namely multi-scattering reactions, and kaon
absorption by one nucleon, $K^- N \to \pi \Sigma$ or $K^- N \to \pi \Lambda$
or by a pair of nucleons, $\bar{K} N N\to  \Sigma N$
and $\bar{K} N N\to  \Lambda N$, followed by decay of $\Sigma$ or $\Lambda$ into $\pi N$.
The contributions from these processes were also suggested in Ref. \cite{YH}.

In the present work, we take
into account all the above mentioned reactions by means
 of a Monte Carlo simulation \cite{magas-inflight}.
As in the experiment \cite{Kishimoto:2007zz},  we select "good events" with fast protons  that emerge
within an angle of 4.1 degrees in the nuclear rest frame (lab frame). 
We plot our obtained $^{12}$C$(K^-,p)$ spectrum as a function of a binding energy of the kaon, $E_B$, should
the process correspond to the trapping of a kaon in a bound state and emission of the fast proton.

%

\begin{figure}[htb]
\vspace{-0.25cm}
\includegraphics[width=.5\textwidth]{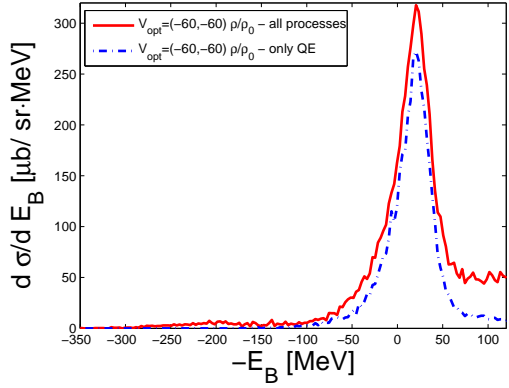}
\vspace{-0.75cm}
\caption{Calculated $ ^{12}C(K^-,p)$ spectra 
 with  $V_{\rm opt}=(-60,-60)\rho/\rho_0$ MeV,
  taking into account only quasi-elastic processes (dash-dotted line),  
  and including all the contributing processes (full line).}
\label{fig1}
\end{figure}

We take into account a kaon optical potential $V_{\rm opt}={\rm Re}\, V_{\rm
opt} + {\rm i}~ {\rm Im}\, V_{\rm opt} $, which will influence the kaon propagation
through the nucleus, especially when it will acquire a relatively low momentum 
after a high momentum transfer quasi-elastic collision.  
In the present study we take the strength of the potential as predicted by chiral models:
${\rm Re}\, V_{\rm opt}= -60\, \rho/\rho_0$ MeV \cite{lutz,angelsself,schaffner,galself,Tolos:2006ny}; 
${\rm Im}\, V_{\rm opt} \approx -60\, \rho/\rho_0$ 
MeV, as in
the experimental paper \cite{Kishimoto:2007zz} and the theoretical study of
\cite{angelsself}.

In the Monte Carlo simulation \cite{magas-inflight} we implement this distribution by
generating a random kaon mass $\tilde{M}_K$ around a central value, 
$M_K + {\rm Re}\,V_{\rm opt}$, within 
a certain extension determined by the width of the distribution
$\Gamma_K = -2 {\rm Im}\, V_{\rm opt}$. The probability assigned to each value
of $\tilde{M}_K$ follows the Breit-Wigner distribution given by the kaon 
spectral function:
\begin{center}
$
S(\tilde{M}_K )=\frac{1}{\pi} 
\frac{-2M_K {\rm Im}\,V_{\rm opt}}
{(\tilde{M}^2_K -M_K^2-2 M_K {\rm Re}\,V_{\rm opt})^2 + (2M_K {\rm Im}\,V_{\rm opt})^2}
\,.
$
\end{center}

In Fig.~\ref{fig1}
we show  the results
of the Monte Carlo simulation obtained with an optical potential
$V_{\rm opt}=(-60,-60)\rho/\rho_0$ MeV \cite{magas-inflight}:  first, taking into account only quasi-elastic
processes; and then taking into account all the discussed mechanisms. We can see that there is some 
strength gained in the region of "bound kaons" due to the new mechanisms.
Although not shown separately in the figure, we have observed
that one nucleon absorption and several rescatterings
contribute to the region $-E_B > -50$ MeV. To some extent, this strength
can be simulated by the parametric background used in
\cite{Kishimoto:2007zz}. However, this is not true anymore for the two nucleon absorption process,
which contributes to all values of $-E_B$, starting from almost as low as $-300$ MeV.

It is very important to keep in mind that in the spectrum of \cite{Kishimoto:2007zz} 
the outgoing forward protons were measured in  coincidence  
with at least one charged particle in the decay counters sandwiching the target.
Obviously, the real simulation of such a coincidence experiment is tremendously
difficult, practically impossible with high accuracy, because it
would require tracing out all the charged particles coming out from all
possible scatterings and decays. 
Although we are studying many processes and 
following many particles in our
Monte Carlo simulation, which is not the case in the Green function method
used in the data analysis \cite{Kishimoto:2007zz}, 
we cannot simulate precisely the real coincidence effect.

\begin{figure}[htb]
\vspace{-0.25cm}
\includegraphics[width=.5\textwidth]{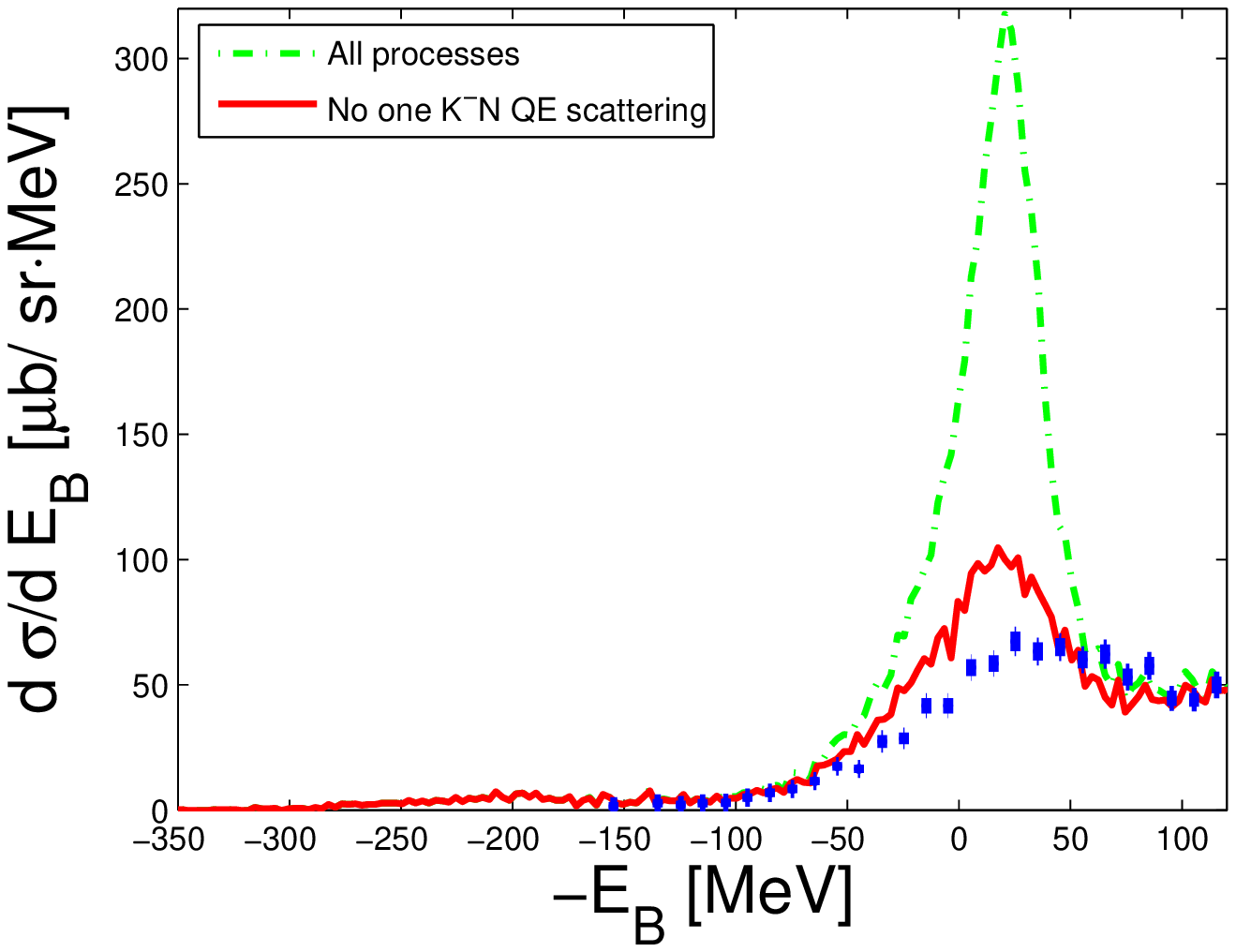}
\includegraphics[width=.5\textwidth]{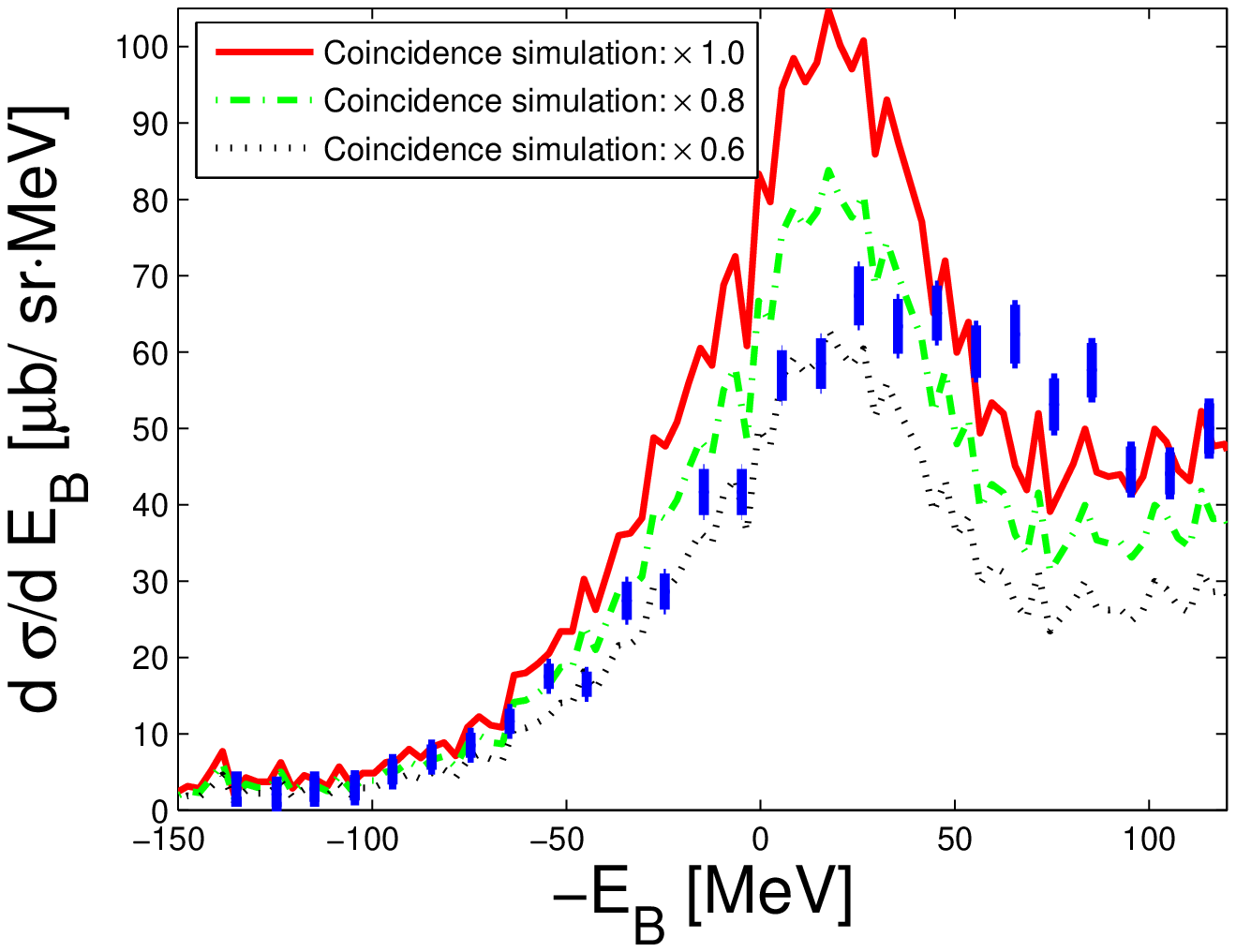}
\vspace{-0.75cm}
\caption{Calculated $ ^{12}C(K^-,p)$ spectra 
 with  $V_{\rm opt}=(-60,-60)\rho/\rho_0$ MeV to be compared with data points from \cite{Kishimoto:2007zz}.
  {\it Left plot: } Dash-dotted line shows the spectra calculated taking into account all contributing processes; then imposing minimal 
  coincidence requirement we obtain spectra shown by a full line. 
   {\it Right plot:} Solid line shows spectra with minimal coincidence requirement, as on the left plot; dash-dotted and dotted lines - 
 the same spectra with additional suppression factors - dash-dotted and dotted lines.  
   }
\label{fig2}
\end{figure}

The best we can do is to eliminate the processes
which, for sure, will not produce a coincidence; this can be 
called minimal coincidence requirement \cite{magas-inflight}.
If the kaon in the first
quasi-elastic scattering produces an energetic proton falling into the
peaked region of the spectra, then the emerging kaon will be 
scattered backwards.
In our Monte Carlo simulations we can select events were neither the proton,
nor the kaon will have any further reaction after such a scattering. In these
cases, although there is a "good" outgoing proton, there are no charged
particles going out with the right angle with respect to the beam axis to
hit a decay counter, since the $K^-$ escapes undetected in the backward
direction. Therefore, this type of events must be eliminated for 
comparison with the experimental spectra.

It is clear from Fig.~\ref{fig1} that the main source of the energetic protons for $ ^{12}C(K^-,p)$ spectra is
$K^-p$ quasi-elastic scattering, however many of these events will not pass the coincidence condition.
Implementing the minimal coincidence requirement, as discussed above, 
we will cut off a substantial part of the potentially "good" events, and drastically
change the form of the final spectrum \cite{magas-inflight}, as illustrated in Fig. \ref{fig2}, left plot.

To further simulate the coincidence requirement we introduce additional constant suppression factors 
to the obtained spectrum - see Fig. \ref{fig2}, right plot. Comparing our results with the experimental data we can 
conclude that in the "bound" region, $-E_B < 0$ MeV, this additional suppression 
is about $\sim 0.7$  and more or less homogeneous, while in the  continuum the suppression weakens and for  
$-E_B > 50$ MeV it is negligible. This picture is natural from the physical point of view, because the 
r.h.s. of the spectrum, Fig. \ref{fig2},  with relatively low momentum protons is mostly populated by many particle final states, 
which have a good chance to score the coincidence. 

To conclude, the main point of our analysis is not to state that
the data of Ref.~ \cite{Kishimoto:2007zz} supports  
${\rm Re}\, V_{\rm opt}=-60\rho/\rho_0$ MeV
rather than $-200\rho/\rho_0$. We want to make it clear that trying to simulate
these data one necessarily introduces large uncertainties due to the
experimental set up.   
Thus, this experiment is not appropriate
 for extracting information on the kaon optical potential \cite{magas-inflight}. 

Contrary to what it is assumed in Ref.~\cite{Kishimoto:2007zz},
we clearly see, Fig. \ref{fig2}, that the spectrum shape is 
affected by the required coincidence. 
The experimental data without the coincidence requirement 
would be a more useful observable.\\


{\bf Acknowledgments.}\ \ \ This work is partly supported by
the contracts FIS2008-01661 from MICINN
(Spain), by CSIC and JSPS under the Spain-Japan research Cooperative program,
 and by the Ge\-ne\-ra\-li\-tat de Catalunya contract 2009SGR-1289. We
acknowledge the support of the European Community-Research Infrastructure
Integrating Activity ``Study of Strongly Interacting Matter'' (HadronPhysics2,
Grant Agreement n. 227431) under the Seventh Framework Programme of EU.



\bibliographystyle{aipproc}   

\end{document}